%% file: main.tex
\documentclass[10pt, a4paper]{article}

\usepackage{lrec-coling2024}
\usepackage{multirow}
\usepackage{multicol}
\usepackage{booktabs}
\usepackage{caption}
\usepackage{subcaption}
\usepackage{floatrow}
\newfloatcommand{capbtabbox}{table}[][\FBwidth]

\usepackage{amsmath,amsfonts}
\title{Modal-adaptive Knowledge-enhanced Graph-based Financial Prediction from Monetary Policy Conference Calls with LLM}

\name{Kun Ouyang$^1$, Yi Liu$^1$, Shicheng Li$^1$, Ruihan Bao$^2$, Keiko Harimoto$^2$, Xu Sun$^1$} 

\address{$^1$National Key Laboratory for Multimedia Information Processing,\\ School of Computer Science, Peking University\\
$^2$Mizuho Securities Co., Ltd. \\
kunouyang10@gmail.com, \{imliuyi, lisc99\}@pku.edu.cn, \\       \{ruihan.bao, keiko.harimoto\}@mizuho-sc.com, xusun@pku.edu.cn\\}

\abstract{
Financial prediction from Monetary Policy Conference (MPC) calls is a new yet challenging task, which targets at predicting the price movement and volatility for specific financial assets by analyzing multimodal information including text, video, and audio. 
Although the existing work has achieved great success using cross-modal transformer blocks, it overlooks the potential external financial knowledge, the varying contributions of different modalities to financial prediction, as well as the innate relations among different financial assets. To tackle these limitations, we propose a novel \textbf{M}odal-\textbf{A}daptive k\textbf{N}owledge-enh\textbf{A}nced \textbf{G}raph-bas\textbf{E}d financial p\textbf{R}ediction scheme, named MANAGER. Specifically, MANAGER resorts to FinDKG to obtain the external related knowledge for the input text. Meanwhile, MANAGER adopts BEiT-3 and Hidden-unit BERT (HuBERT) to extract the video and audio features, respectively. Thereafter, MANAGER introduces a novel knowledge-enhanced cross-modal graph that fully characterizes the semantic relations among text, external knowledge, video and audio, to adaptively utilize the information in different modalities, with ChatGLM2 as the backbone. Extensive experiments on a publicly available dataset Monopoly verify the superiority of our model over cutting-edge methods.
 \\ \newline  }

\begin{document}

\maketitleabstract

\input{Chapter/1_introduction.tex}

\input{Chapter/2_related.tex}
\input{Chapter/3_method.tex}
\input{Chapter/4_experiment.tex}
\input{Chapter/5_conclusion.tex}
\section{Acknowledgement}
We thank all the anonymous reviewers for their valuable
suggestions.
This work is supported by a Research Grant from Mizuho Securities Co., Ltd. We sincerely thank Mizuho Securities for valuable domain expert suggestions. Ruihan Bao and Xu Sun are the corresponding authors.

\nocite{*}
\section{Bibliographical References}\label{sec:reference}
\bibliographystyle{lrec-coling2024-natbib}
\bibliography{reference}

\bibliographystylelanguageresource{lrec-coling2024-natbib}
\bibliographylanguageresource{languageresource}

\end{document}

%% file: Chapter/1_introduction.tex
\section{Introduction}
Forecasting the fluctuation of prices for a financial asset over a specific period is a crucial task in financial analysis, essential for both investors and policymakers~\cite{Lewellen2003FinancingDW}. Accurate prediction results can assist investors in making informed decisions regarding investment returns, while policymakers can implement prudent monetary measures to uphold a robust economy~\cite{Cai2021ItsNA,Shapiro2019TakingTF}. 
In early work, researchers made efforts to solve financial prediction for textual financial data, such as BloombergGPT~\cite{Wu2023BloombergGPTAL} and FinGPT~\cite{wang2023fingptbenchmark}. 
\begin{figure}[h]
    \centering
    \includegraphics[scale=0.44]{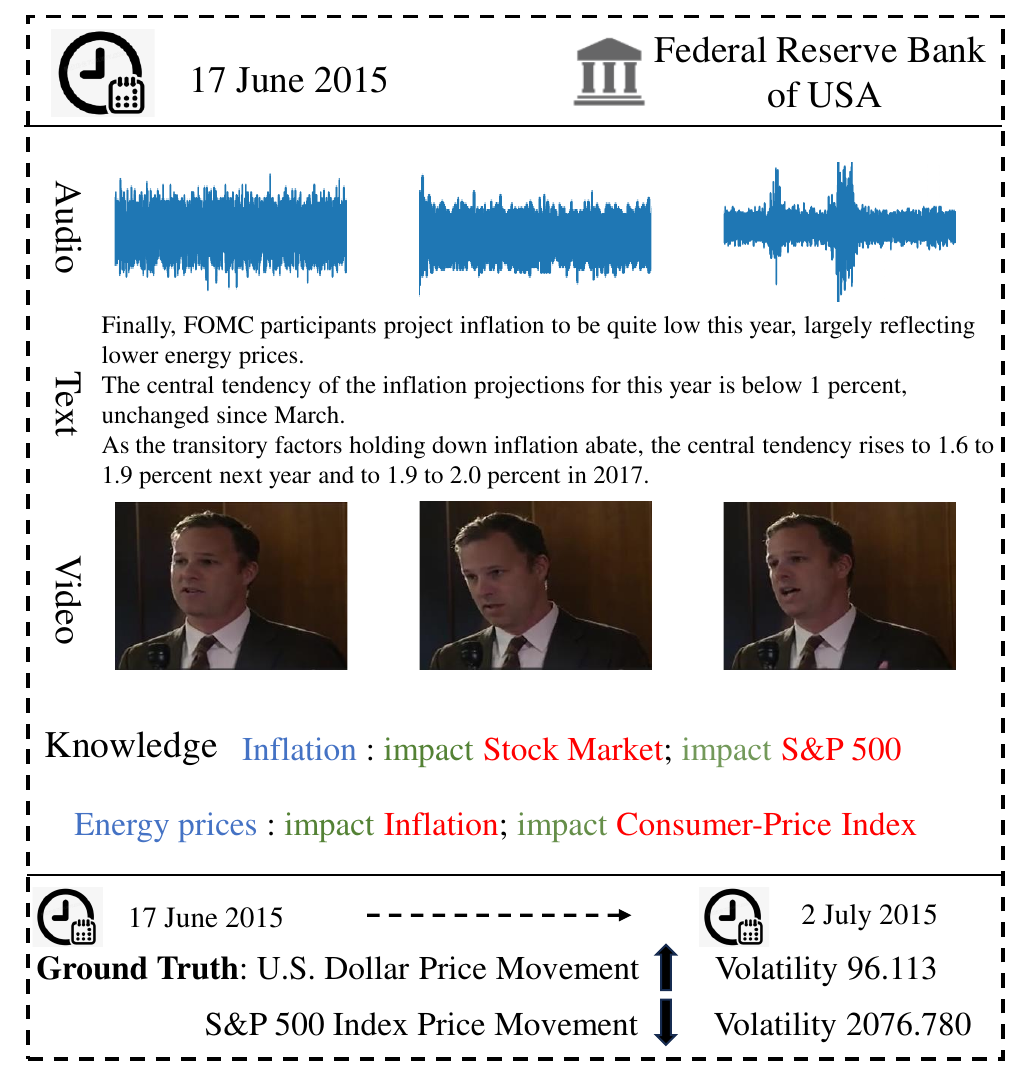}
    \caption{An example of the financial prediction from MPC calls. We also present the external knowledge inferred by FinDKG for the given text. Notably, the words in blue are the anchor entities while those in green are the relations and those in red are the related entities.}
    \label{fig:intro}
    \vspace{-0.2cm}
\end{figure}

Despite their promising performance, the above models can only solve text-based financial tasks.
With unprecedented advances in multimodal learning, investors now have access to a vast amount of unstructured data for financial prediction~\cite{Jiang2020ApplicationsOD}. Moreover, the non-verbal information involved in the visual and acoustical modalities (e.g., vocal tone and facial expressions) can be indicative and correlated with trading activities in the financial market. 
One such abundant source of multimodal information is the Monetary Policy Conference (MPC's) call. Previous work~\cite{Boukus2006TheIC} has underscored the influence of MPC calls on financial stock markets. Therefore, \citeauthor{Mathur2022MONOPOLYFP}~(\citeyear{Mathur2022MONOPOLYFP}) curated a public Monetary Policy Call
Dataset named Monopoly and proposed to predict the price movement and volatility for six principal financial assets (i.e., Stock Index (Small), Stock Index (Large), Gold Price,
Currency Exchange Rate, Long-term bond yield (10-years), Short-term
bond yield (3-months)) based on MPC calls. The authors adopted cross-modal transformer blocks and
modality-specific attention fusion to conduct price movement and volatility prediction.
Although this pioneering study has achieved promising performance, it still suffers from three key limitations.

1) \textbf{Overlook the potential external knowledge in the financial domain.}
The pioneering study fails to utilize the related knowledge contained in the external public knowledge base in the financial domain. As shown in Figure~\ref{fig:intro}, the related knowledge obtained from FinDKG~\cite{1} can strengthen the context comprehension (e.g., ``impact S\&P 500'') and promote the financial prediction. 

2) \textbf{Overlook the varying contributions of different modalities to financial prediction.}
The existing work equally feeds the multimodal features (i.e., text, video and audio) into the model, and treats them as the equal source of information to conduct multimodal information fusion with the same weights. In fact, the content of given text is the prime cue for the financial prediction, while the non-verbal cues such as facial expressions and vocal tone involved in the video and audio play a minor role in comprehending the context. How to adaptively utilize the information residing in the multiple modalities merits our attention.

3) \textbf{Overlook the innate relations among different financial assets.}
The former method predicts the price movement and volatility of six financial assets independently, ignoring the potential relationships among different financial assets. Actually, the price changes of a financial asset may provide useful information to predict price trend of the other financial assets.

\begin{figure*}
    \centering
    \includegraphics[scale=0.45]{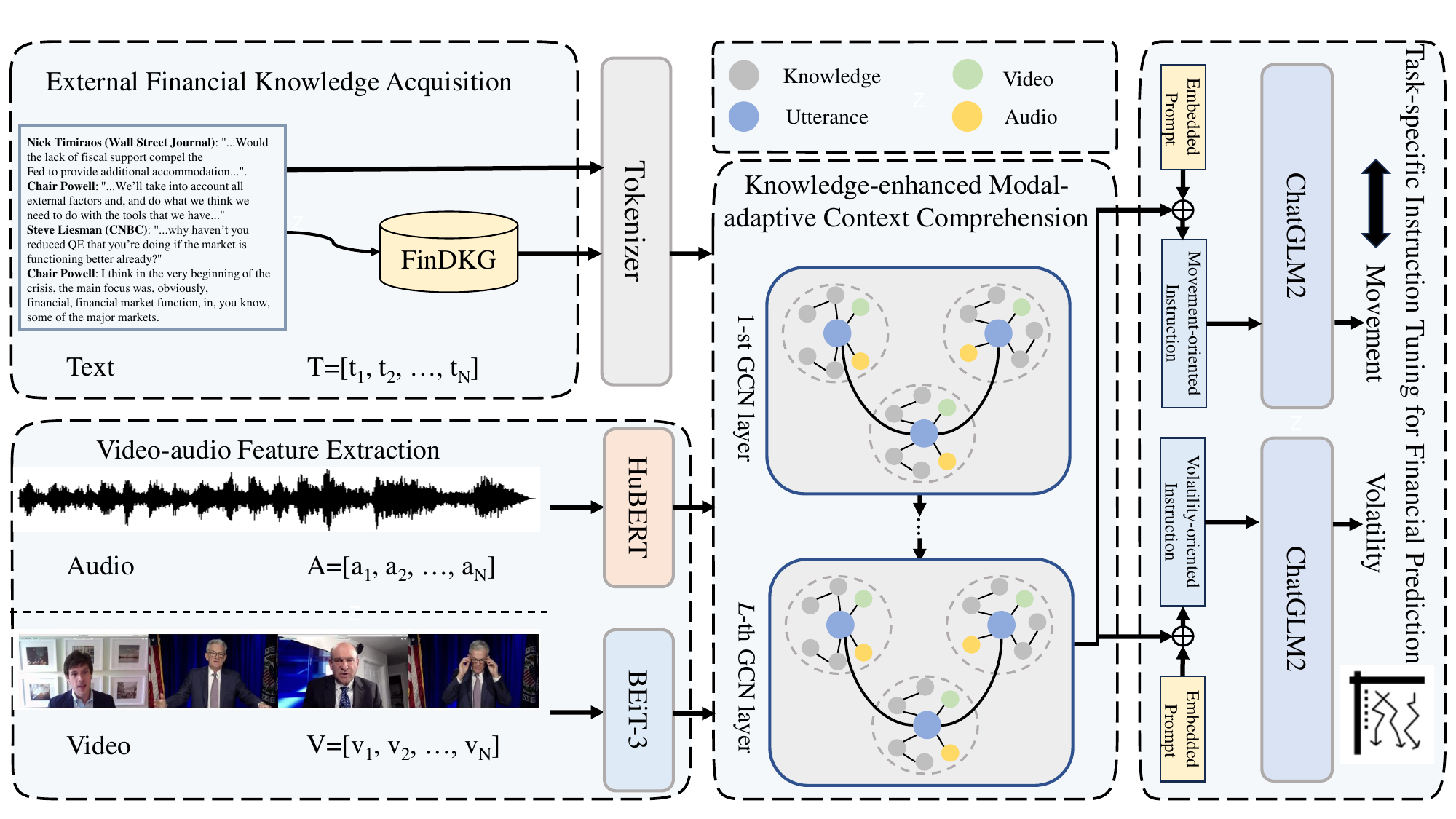}
    \caption{The architecture of MANAGER, which consists of four key components including External Financial Knowledge Acquisition, Video-audio Feature Extraction, Knowledge-enhanced Modal-adaptive Context Comprehension and Task-specific Instruction Tuning for Financial Prediction.}
    \label{fig:framework}
\end{figure*}

To tackle these limitations, we propose a novel \textbf{M}odal-\textbf{A}daptive k\textbf{N}owledge-enh\textbf{A}nced \textbf{G}raph-bas\textbf{E}d financial p\textbf{R}ediction scheme, MANAGER for short.
In detail, MANAGER consists of four components: external financial knowledge acquisition, video-audio feature extraction, knowledge-enhanced modal-adaptive context comprehension and task-specific instruction tuning for financial prediction, as shown in Figure~\ref{fig:framework}. 
In the first module, we focus on acquiring the external related knowledge for the given text, where a large-scale financial knowledge base FinDKG~\cite{1} is used. In the second module, we utilize BEiT-3~\cite{DBLP:journals/corr/abs-2208-10442} and Hidden-Unit BERT (HuBERT)~\cite{Hsu2021HuBERTSS} to extract the video and audio representations, respectively. In the third module, we construct the knowledge-enhanced cross-modal graph to aggregate the given text, input video, audio and inferred external knowledge through two types of relations (i.e., intra-modal and inter-modal semantic relations). We then employ the commonly used graph neural networks (GCNs)~\cite{DBLP:conf/iclr/KipfW17}, which have shown great performance in NLP tasks~\cite{Jing2023MultisourceSG,DBLP:journals/corr/abs-2402-03658}, to adaptively utilize the different modalities for cross-modal context comprehension. In the last module, considering that up-to-date Large Language Models (LLMs) have shown promising performance in multimodal context learning~\cite{Zhang2023CanLL,Wu2023AutoGenEN}, the potential of LLMs in the multimodal financial prediction task is increasingly evident.
Therefore, we adopt ChatGLM2~\cite{du2022glm} as the backbone and feed the cross-modal representation into ChatGLM2 with a task-specific instruction devised for the certain financial asset to predict the price movement or volatility, respectively. Unlike previous work, we do not conduct prediction for different financial assets independently, but utilize ChatGLM2 to capture the innate relation among the financial assets. Finally, we conduct extensive experiments on a publicly available Monopoly dataset, on which our method outperforms the best baseline across all the metrics for both price movement and volatility prediction. 
Our contributions can be concluded as follows.
\begin{itemize}
    \item We propose a novel modal-adaptive knowledge-enhanced graph-based financial prediction scheme, where the text, external knowledge, video and audio are aggregated for cross-modal context comprehension.
    \item As far as we know, we are the first to introduce an up-to-date LLM named ChatGLM2 to solve the financial prediction task for Monetary Policy Conference (MPC) calls, which contain multiple modalities (i.e., text, video and audio). 
    \item 
    The results of extensive experiments on the Monopoly dataset demonstrate the superiority of our MANAGER over other cutting-edge methods, and prove the effectiveness of each component of MANAGER.
    As a byproduct, we release our code and parameters\footnote{\url{https://github.com/OuyangKun10/MANAGER}.} to facilitate the research community.
\end{itemize}

%% file: Chapter/2_related.tex
\section{Related Work}
\subsection{Large Language Models (LLMs) in Finance}
In early work about the application of LLMs in Finance, researchers resorted to BERT~\cite{DBLP:conf/naacl/DevlinCLT19} to conduct financial tasks, such as FinBert~\cite{ijcai2020p622}, which is dedicated for financial sentiment analysis with under one billion parameters, fine-tuned on a rich financial corpus to excel in finance-specific tasks. Although it achieves promising performance, it falls short of comprehending the long and complex financial text.
In recent years, there has been a surge in research dedicated to integrating financial datasets with GPT-based models~\cite{Brown2020LanguageMA}, 
aimed at enhancing Natural Language Processing (NLP) applications.
For example, BloombergGPT~\cite{Wu2023BloombergGPTAL} is a closed-source model, trained extensively on
diverse financial datasets, thereby encapsulating a broad spectrum of the financial domain.
FinGPT~\cite{wang2023fingptbenchmark} is an open-source LLM, fine-tuned from a general LLM using low-rank
adaptation method~\cite{Hu2021LoRALA}, fostering accessibility for the broader community.


\subsection{Multimodal Financial Prediction}
Existing work in the financial realm utilize vocal and textual cues from earnings conference calls~\cite{Qin2019WhatYS,Sawhney2020MultimodalMF}, and mergers and acquisitions calls~\cite{Sawhney2021MultimodalMM} for stock volatility prediction. Multimodal architectures that use these
cues for financial predictions have seen significant improvements in their performances~\cite{Sawhney2020MultimodalMF,Yang2020HTMLHT}. However, the vision modality, which
may offer important cues that correlate with the performance of financial markets~\cite{Cao2021AIIF} remains underexplored. Therefore, \citeauthor{Mathur2022MONOPOLYFP} (\citeyear{Mathur2022MONOPOLYFP}) first introduced video modality in the financial prediction task and released a dataset named Monopoly. They adopted cross-modal transformer blocks and
modality-specific attention fusion to forecast the financial risk and price movement. Despite its promising performance on financial prediction, it overlooks the potential external knowledge, the varying contributions of different modalities, the innate relations among different financial assets, which are the major concerns of our model.

%% file: Chapter/3_method.tex
\section{Task Formulation}
Suppose we have a training dataset $\mathcal{D}$ composed of $N_d$ samples, i.e., $\mathcal{D}=\{d_1,d_2,\cdots,d_{N_d}\}$. Each sample $d_i=\{T_i,V_i,A_i,Y_i\}$, where $T_i=\{u_1^i, u_2^i, \cdots u_N^i\}$ denotes the input text containing $N$ utterances, $V_i=\{v_1^i, v_2^i, \cdots v_N^i\}$ and $A_i=\{a_1^i, a_2^i, \cdots a_N^i\}$ are the set of the input video and audio clips, respectively. 
Each utterance $u_i$ contains $N_{u_i}$ tokens. i.e., $u_i=\{t_1^i,t_2^i,\cdots,t_{N_{u_i}}^i\}$
And $Y_i^\tau=\{p^\tau_i,o^\tau_i\}$ denotes the target labels over a period of $\tau$ days, where $p^\tau_i$ is the price movement and $o^\tau_i$ is the volatility, respectively. Our target is to learn a multimodal financial prediction model $\mathcal{F}$ that is able to predict the price movement and volatility for six principal financial assets (i.e., Stock Index (Small), Stock Index (Large), Gold Price,
Currency Exchange Rate, Long-term bond yield (10-years), Short-term
bond yield (3-months) ), based on the given multimodal input as follows,
\begin{equation}
    \hat{Y}_i = \mathcal{F}(T_i,V_i,A_i|\Theta)
\end{equation}
where $\Theta$ is a set of learnable parameters of the model $\mathcal{F}$. $\hat{Y}_i=\{\hat{p}^\tau_i,\hat{o}^\tau_i\}$ is the labels (i.e., price movement and volatility) predicted by 
 $\mathcal{F}$. For simplicity, we omit the subscript $i$ that indexes the training samples.
 
\section{Method}
In this section, we detail the four components of our proposed MANAGER, as shown in Figure~\ref{fig:framework}.

\subsection{External Financial Knowledge Acquisition}
As aforementioned, the external financial knowledge inferred by the input text can assist the financial prediction since it can introduce corresponding financial entities as well as the relations, and provide some external factors to analyze the financial environment, leading to more informed predictions. Specifically, we resort to FinDKG~\cite{1}, which provides dynamic knowledge graph data in the financial domain, as the source of external knowledge. Notably, FinDKG changes dynamically over time. In detail, it contains $13,645$ financial entities and $15$ types of relations. Given the input text, we adopt the period-specific FinDKG\footnote{\url{https://xiaohui-victor-li.github.io/FinDKG/}.} that only contains the knowledge before the date of the input text, to prevent our model from obtaining the information beyond the date. The ration is that the information beyond the date can influence the prediction. 

To acquire the related external knowledge for the given text, i.e., $T=\{u_1, u_2, \cdots u_N\}$, we first identify all the entities in FinDKG that exist in the input text. Let $\{e_1, \cdots, e_{N_e}\}$ be the set of identified entities, where $N_e$ is the total number of the identified entities. We then use these identified entities as the anchors to obtain the related entities and corresponding relations as the external knowledge for the input text. Notably, for each anchor entity $e$, we retrieve all its one-hop neighboring entities, as well as the corresponding relations that are treated as the edges, from FinDKG and deem them as the external knowledge for $e$. 
Mathematically, let $\mathcal{N}(e)=\mathcal{N}^1(e) \cup \mathcal{N}^2(r)$ be the set of external knowledge (i.e., $\mathcal{N}^1(e)$ is the set of neighboring entities and $\mathcal{N}^2(r)$ is the set of corresponding relations between each neighboring entity and the anchor entity, respectively.) of the entity $e$ in FinDKG. Then the related external knowledge for the input text can be represented as $\{\mathcal{N}^1_{e_1},\mathcal{N}^1_{e_2}, \cdots, \mathcal{N}^1_{e_{N_e}}\}\cup\{\mathcal{N}^2_{e_1},\mathcal{N}^2_{e_2}, \cdots, \mathcal{N}^2_{e_{N_e}}\}$. $N_e$ is the number of the neighboring entities as well as the number of the relations.
\subsection{Video-audio Feature Extraction}
To obtain the global feature of video and audio clips, we choose BEiT-3~\cite{DBLP:journals/corr/abs-2208-10442} and Hidden-Unit BERT (HuBERT)~\cite{Hsu2021HuBERTSS} as the visual and acoustical encoder, respectively.

\textbf{Video Encoding},
to encode the video clips, we resort to BEiT-3, which is an advanced general-purpose multimodal foundation model pre-training for all vision and vision-language tasks and shows great performance in visual modality encoding~\cite{DBLP:journals/corr/abs-2208-10442}. Specifically, we embed each frame $v^k_j$ in the video clip $v_j$ as the arithmetic mean of visual tokens representations of that frame. We then average over all the frames to obtain the aggregated encoding feature $x_V^j \in \mathbb{R}^{D}$, where $D$ is the feature dimension. Mathematically, we have
\begin{equation}
  x_V^j=\frac{1}{N_f}\sum_{k=1}^{N_f} \operatorname{BEiT-3}(v^k_j), \forall j \in [1,N],\label{eq1}
\end{equation}
where $N_f$ is the number of frames in the clips $v_j$. And we represent the sequence
of video features as $X_V=[x_v^1,x_v^2,\cdots, x_v^{N}]$.

\textbf{Audio Encoding}, we extract the feature of the audio clips via the self-supervised speech representation model named HuBERT, which has shown significant power for extracting
audio features for speech language understanding tasks~\cite{Yoon2022HuBERTEEEE}. We embed each audio utterance $a^k_j$ in the audio clip $a_j$ as the arithmetic mean of the representation derived by HuBERT and obtain the encoded acoustical feature $x_A^j \in \mathbb{R}^{D}$. Formally,
\begin{equation}
  x_A^j=\operatorname{HuBERT}(a^k_j), \forall j \in [1,N],\label{eq2}
\end{equation}
we represent the sequence
of audio features as $X_A=[x_a^1,x_a^2,\cdots, x_a^{N}]$.
\subsection{Knowledge-enhanced Modal-adaptive Context Comprehension}
In this module, we aim to enhance the cross-modal context comprehension utilizing the inferred external knowledge in the financial domain. In fact, there are rich relations (i.e., intra-modal semantic relation and inter-modal semantic relation) existing in the multiple input including the text, video, audio and external knowledge. Therefore, to adaptively utilize different modalities via these semantic relations for boosting cross-modal context comprehension, we resort to the widely used graph neural networks (GCNs). Specifically, we first build a knowledge-enhanced cross-modal graph $\mathcal{G}$.

\subsubsection{Nodes Initialization} In particular, the nodes in the knowledge-enhanced cross-modal graph $\mathcal{G}$ come from four kinds of sources, the given text $T$, input video clips $V$, input audio clips $A$ and inferred external knowledge $\mathcal{N}(e)$. We define all the nodes as $\{n_1,\cdots,n_{N_c}\}=\{T, \mathcal{N}(e), V, A\}$, where $N_c$ is the total number of nodes. To initialize the nodes, we feed the textual input $\{T, \mathcal{N}(e)\}$ into the encoder of ChatGLM2~\cite{du2022glm} to extract their features. Specifically, we first concatenate them into a sequence of tokens, denoted as $X_T=\{T, \mathcal{N}(e)\}$, and then feed $X$ into the encoder $\mathcal{E}$ as follows,
\begin{equation}
   \mathbf{H}=\mathcal{E}(X_T),\label{eq3}
\end{equation} 
where $\mathbf{H}=[\mathbf{h}_1, \cdots, \mathbf{h}_{N_t+2 \times N_e}] \in \mathbb{R}^{(N_t+2 \times N_e) \times D}$ is the encoded representation matrix, $N_t$ is the tokens number of the whole utterances and each column of which corresponds to a token. Accordingly, nodes of the textual part (utterances and external knowledge) in the knowledge-enhanced cross-modal graph $\mathcal{G}$ can be initialized by $\mathbf{H}$, where the $j$-th token node is initialized with $\mathbf{h}_j$.
In addition, the other nodes are initialized by the extracted video feature sequence $X_V$ and the extracted audio feature sequence $X_A$, respectively.

\subsubsection{Semantic Relation Construction}
To enhance the cross-modal context comprehension with related external knowledge, we consider two kinds of semantic relations: intra-modal semantic relation and inter-modal semantic relation. The former captures the basic information flow of the multiple modalities input, also incorporates the related external knowledge into the text. The latter enables the injection of non-textual information from video and audio into the context and achieves cross-modal information fusion.

\textbf{Intra-modal Semantic Relation}.
To capture the information flow in the specific modality, we design three types of intra-modal semantic edges.
a) \textit{Token-token edges}. We introduce an edge between each pair of adjacent nodes in given text to represent the neighboring relations among the tokens of text. b) \textit{Token-knowledge edge}. We connect the tokens that act as an anchor entity in the aforementioned external knowledge retrieval process, relation token and the related entity token sequentially.
c) \textit{Video-video edge} and d) \textit{Audio-audio edge}. We link each pair of adjacent video nodes and connect each pair of adjacent audio nodes to represent the adjacency relations of the video and audio modalities, respectively.
The above edges are characteristics of the information flow, and weighted by $1$. Formally, we introduce the corresponding adjacency matrix $\mathbf{A}^1$ for representing these edges as follows,
\begin{equation}
 \mathbf{A}^1_{i,j}=\left\{
 \begin{aligned}
&1,\quad if~{D_1}(n_i,n_j), \\\
&0,\quad otherwise,
\end{aligned}\label{eq:graph}
 \right.
\end{equation}
where $N_c$ denotes the total number of nodes in $\mathcal{G}$ and $i,j \in [1, N_c]$. $D_1(n_i,n_j)$ denotes that the nodes $n_i$ and  $n_j$ have the certain above defined intra-modal semantic relation. 

\textbf{Inter-modal Semantic Relation}.
To comprehensively utilize the multiple modalities to promote cross-modal context comprehension, we devise two types of inter-modal semantic edges.
a) \textit{Token-video edges}. For each video node, we connect it to each token in the corresponding utterance. The ration is to inject the visual information (e.g., facial expressions and hand gestures) that can help semantics understanding and hence improve financial analysis, into the context.
b) \textit{Token-audio edges}. For each audio node, we link it with each token in the corresponding utterance. In this way, we can incorporate the acoustic information (e.g., vocal tone) that is also useful for context comprehension, into the context. The weight of all the above edges is set to $1$.
Accordingly, the adjacency matrix $\mathbf{A}^2 \in \mathbb{R}^{N_c \times N_c}$ for capturing the above inter-modal semantic relations can be constructed as follows, 
\begin{equation}
 \mathbf{A}^2_{i,j}=\left\{
 \begin{aligned}
1 & ,\quad if~D_2(n_i,n_j), \\\
0 & ,\quad otherwise,
\end{aligned}\label{eq:graph_1}
 \right.
\end{equation}
where $D_2(n_i, n_j)$ indicates that nodes $n_i$ and $n_j$ have certain above inter-modal semantic semantic relation, $i\in [1, N_t]$ and $j\in [N_t+2 \times N_e+1, N_c]$. 

Ultimately, by combing the adjacency matrices for intra-modal and inter-modal semantic relations, i.e., $\mathbf{A}^1$ and $\mathbf{A}^2$, we can derive the final adjacency matrix $\mathbf{A}$ for the knowledge-enhanced cross-modal graph. 

\subsubsection{Graph Convolution Network}
Towards the final cross-modal context comprehension, we adopt $L$ layers of GCN to extract the multimodal fusion feature of the cross-modal context. Then the node representations are iteratively updated as follows,
\begin{equation}
    \mathbf{G}_{l}=ReLU(\tilde{\mathbf{A}}\mathbf{G}_{l-1}\mathbf{W}_l), l \in [1,L],
\end{equation}
where $\tilde{\mathbf{A}} = (\mathbf{D})^{-\frac{1}{2}}\mathbf{A}(\mathbf{D})^{-\frac{1}{2}}$ is the normalized symmetric adjacency matrix, and $\mathbf{D}$ is the degree matrix of the adjacency matrix $\mathbf{A}$. In addition, $\mathbf{W}_l \in \mathbb{R}^{D \times D}$ is a trainable parameter of the $l$-th GCN layer. 
$\mathbf{G}_l$ are the node representations obtained by the $l$-th GCN layer, where 
$\mathbf{G}_0=\mathbf{H}$ is the initial node representation. 
\input{Table/main_results}
\subsection{Task-specific Instruction Tuning for Financial Prediction}
The final nodes representation $\mathbf{G}_L$ obtained by the $L$-th layer GCNs absorb rich semantic information from their correlated nodes and can be used as the input for the following financial prediction. Considering that we need to solve a couple of tasks (i.e., prediction of the price movement and volatility), we resort to the advanced large language model ChatGLM2, which shows great performance in context comprehension~\cite{du2022glm}, and fine-tune it for each task independently. In addition, constructing proper instructions is pivotal for task-specific tuning, with each task being guided by a unique instruction prompt. Therefore, we adopt the instruction template~\cite{Wang2023FinGPTIT} structured as follows:

\texttt{Instruction}: $[prompt]$ Input: $[input]$ Answer: $[output]$

This template provides a standardized format, facilitating consistency across different tasks.
Notably, we utilize the aforementioned final nodes representation $\mathbf{G}_L$ as input.
Next, we design prompt for specific task.

\textbf{Movement Prediction}.
In this task, we aim to predict the price movement for the financial assets. Therefore, the movement-oriented prompt is designed to guide ChatGLM2 to judge the price movement (e.g., ``increase'' or ``decrease'') of the given asset based on the multimodal input. The prompt template is ``Please predict the price movement of $O$ in $\tau$ days after the $date$ according to the input'', where $O$ is the to-be-predicted financial asset, $\tau \in \{1,3,7,15\}$ and $date$ is formatted as \texttt{YYYY-MM-DD}. 

\textbf{Volatility Prediction}. 
In this task, we aim to predict the volatility, a float number that measures the instability of an asset. Therefore, the volatility-oriented prompt is designed to guide ChatGLM2 to output the volatility of the given financial asset based on the multimodal input. Similar to the above prompt template, we just replace ``price movement'' with ``volatility''. 

We then utilize encoder of ChatGLM2 to embed the prompt and concatenate it with the input. $[output]$ is the prediction result that is answered after we feed the instruction into ChatGLM2. And we can obtain the task-specific instruction (i.e., Movement-oriented instruction $I_p$ and volatility-oriented instruction $I_v$). Finally, we feed $I_p$ and $I_v$ into ChatGLM2 independently to guide it to conduct the two financial prediction tasks. 
For optimizing our MANAGER, we adopt Binary Cross-Entropy (BCE) loss and Mean Squared Error (MSE) loss to train the output for price movement prediction and volatility prediction, respectively.

%% file: Table/main_results.tex
\begin{table*}
\centering
\caption{Performance comparison with baselines for movement prediction in terms of F1 score $\tau$-days after the call ($\tau \in \{1,3,7,15\}$). The best results are in boldface, while the second best are underlined. $\star$ denotes that the p-value of the significant test between our result and the best baseline result is less than 0.01.}
\label{Tab:Comparison movement}
\begin{floatrow}
\begin{subtable}[t]{0.5\linewidth}
\resizebox{\linewidth}{!}{
\begin{tabular}{l|cccc|cccc|cccc}
\toprule
\multirow{2}{*}{\textbf{Model}} & \multicolumn{4}{c|}{\textbf{Stock Index (Small)}}                                                                         & \multicolumn{4}{c|}{\textbf{Stock Index (Large)}} & \multicolumn{4}{c}{\textbf{Currency Exchange Rate}}  \\ 
\cmidrule{2-13} 
& \textbf{F1\_1} & \textbf{F1\_3} & \textbf{F1\_7} & \textbf{F1\_15} & \textbf{F1\_1} & \textbf{F1\_3} & \textbf{F1\_7} & \textbf{F1\_15} & \textbf{F1\_1} & \textbf{F1\_3} & \textbf{F1\_7} & \textbf{F1\_15} \\
\midrule
HistPrice & 0.390 & 0.470 & 0.400 & 0.420 & 0.430 & 0.430 & 0.410 & 0.420 & 0.190 & 0.260 & 0.210 & 0.230 \\
P-SVM & 0.400 & 0.480 & 0.340 & 0.530 & 0.433 & 0.490 & 0.338 & 0.500 & 0.190 & 0.270 & 0.190 & 0.370 \\
P-LSTM &0.410 &0.473 &0.291 &0.546 &0.399 &0.391 &0.421 &0.442 &0.123 &0.232  &0.165 &0.341\\
\midrule
MLP  &0.349 &0.435 &0.209 &0.539  &0.267 &0.319 &0.331 &0.351  &0.101 &0.201 &0.124 &0.311\\
LSTM  &0.449 &0.435 &0.269 &0.527 &0.414 &0.596 &0.371 &0.432 &0.137 &0.229 &0.199 &0.369\\
MMIM  &0.435 &\underline{0.653} &0.302 &0.605 &0.392 &\underline{0.631} &0.329 &0.601 &0.296 &0.217 &0.142 &0.385\\
MDRM  &0.449 &0.419 &0.462 &0.355 &0.409 &0.392 &0.494 &0.324 &0.177 &0.161 &0.379 &0.152\\
HTML  &0.490 &0.645 &0.458 &0.541 &0.431 &0.504 &0.557 &0.482 &0.484 &0.531 &0.298 &\underline{0.626}\\
MULT  &0.491 &0.630 &0.536 &0.629 &0.443 &0.625 &0.572 &0.612 &0.499 &0.547 &\underline{0.473} &0.521\\
MPCNet & \underline{0.501} & 0.590 & \underline{0.565} & \underline{0.638} & \underline{0.460} & 0.590 & \underline{0.559} & \underline{0.620} & \underline{0.520} & \underline{0.570} & 0.329 & 0.450 \\
\midrule
\textbf{MANAGER} & $\textbf{0.548}^{\star}$ & $\textbf{0.694}^{\star}$ & $\textbf{0.610}^{\star}$ & $\textbf{0.659}^{\star}$ &$\textbf{0.517}^{\star}$ & $\textbf{0.652}^{\star}$ & $\textbf{0.589}^{\star}$ & $\textbf{0.646}^{\star}$ & $\textbf{0.564}^{\star}$ & $\textbf{0.608}^{\star}$ & $\textbf{0.511}^{\star}$ & $\textbf{0.681}^{\star}$ \\

\bottomrule
\end{tabular}}
\subcaption{Stock Indices and Currency Exchange Rate}
\end{subtable}

\begin{subtable}[t]{0.5\linewidth}
\resizebox{\linewidth}{!}{
\begin{tabular}{l|cccc|cccc|cccc}
\toprule
\multirow{2}{*}{\textbf{Model}} & \multicolumn{4}{c|}{\textbf{Gold}}                                                                         & \multicolumn{4}{c|}{\textbf{10-Year Bond Yield}} & \multicolumn{4}{c}{\textbf{3-Month Bond Yield}}  \\ 
\cmidrule{2-13} 
& \textbf{F1\_1} & \textbf{F1\_3} & \textbf{F1\_7} & \textbf{F1\_15} & \textbf{F1\_1} & \textbf{F1\_3} & \textbf{F1\_7} & \textbf{F1\_15} & \textbf{F1\_1} & \textbf{F1\_3} & \textbf{F1\_7} & \textbf{F1\_15} \\
\midrule
HistPrice  &0.360 &0.390 &0.350 &0.400 &0.310 &0.290 &0.220 &0.390 &0.220 &0.160 &0.340 &0.330\\
P-SVM  &0.390 &0.420 &0.370 &0.380 &0.340 &0.310 &0.330 &0.330 &0.370 &0.220 &0.310 &0.390\\
P-LSTM  &0.365 &0.352 &0.371 &0.346 &0.320 &0.291 &0.342 &0.258 &0.377 &0.234 &0.332 &0.314\\
\midrule
MLP  &0.243 &0.215 &0.288 &0.315 &0.244 &0.299 &0.234 &0.174 &0.332 &0.157 &0.248 &0.394\\
LSTM  &0.361 &0.337 &0.304 &0.345 &0.364 &0.311 &0.255 &0.394 &0.381 &0.168 &0.382 &0.444\\
MMIM  &0.209 &0.508 &0.412 &0.318 &0.411 &0.318 &0.345 &0.138 &0.417 &0.306 &0.417 &0.379\\
MDRM  &0.434 &0.383 &0.214 &0.317 &0.287 &0.242 &0.314 &0.149 &0.346 &0.198 &\underline{0.478} &0.505\\
HTML  &0.441 &0.654 &0.379 &0.526 &0.529 &0.278 &0.466 &0.389 &0.424 &0.314 &0.397 &0.450\\
MULT  &0.329 &0.590 &\underline{0.454} &0.533 &\underline{0.534} &\underline{0.364} &0.485 &0.400 &0.428 &0.171 &0.466 &0.493\\
MPCNet &\underline{0.444} &\underline{0.668} &0.413 &\underline{0.637} &0.386 &0.327 &\underline{0.560} &\underline{0.625} &\underline{0.493} &\underline{0.556} &0.374 &\underline{0.537} \\
\midrule
\textbf{MANAGER} &$\textbf{0.486}^{\star}$ &$\textbf{0.696}^{\star}$ &$\textbf{0.507}^{\star}$ &$\textbf{0.672}^{\star}$  &$\textbf{0.612}^{\star}$  &$\textbf{0.391}^{\star}$ &$\textbf{0.587}^{\star}$  &$\textbf{0.649}^{\star}$ &$\textbf{0.521}^{\star}$ &$\textbf{0.583}^{\star}$ &$\textbf{0.519}^{\star}$ &$\textbf{0.574}^{\star}$ \\
\bottomrule
\end{tabular}}
\subcaption{Gold Prices, Long-term (10-Years) and Short-term (3-Months) Bonds}
\end{subtable}

\end{floatrow}

\end{table*}

\begin{table*}
\centering
\caption{Performance comparison with baselines for volatility prediction in terms of MSE $\tau$-days after the call ($\tau \in \{1,3,7,15\}$). The best results are in boldface, while the second best are underlined. $\star$ denotes that the p-value of the significant test between our result and the best baseline result is less than 0.01.}
\label{Tab:Comparison volatility}
\begin{floatrow}
\begin{subtable}[t]{0.5\linewidth}
\resizebox{\linewidth}{!}{
\begin{tabular}{l|cccc|cccc|cccc}
\toprule
\multirow{2}{*}{\textbf{Model}} & \multicolumn{4}{c|}{\textbf{Stock Index (Small)}}                                                                         & \multicolumn{4}{c|}{\textbf{Stock Index (Large)}} & \multicolumn{4}{c}{\textbf{Currency Exchange Rate}}  \\ 
\cmidrule{2-13} 
& $\mathbf{MSE_1}$ &$\mathbf{MSE_3}$ & $\mathbf{MSE_7}$ & $\mathbf{MSE_{15}}$ & $\mathbf{MSE_1}$ & $\mathbf{MSE_3}$ & $\mathbf{MSE_7}$ &$\mathbf{MSE_{15}}$ & $\mathbf{MSE_1}$ & $\mathbf{MSE_3}$ & $\mathbf{MSE_7}$ & $\mathbf{MSE_{15}}$ \\
\midrule
HistPrice  &2.486 &2.234 &1.880 &1.664 &3.397 &3.316 &2.934 &2.972 &2.709 &3.187 &3.127 &3.291\\
P-SVM  &2.489 &2.220 &1.915 &1.753 &2.568 &2.921 &1.971 &2.012 &2.104 &2.534 &1.921 &2.231\\
P-LSTM &2.421 &2.217 &1.845 &1.731 &2.128 &2.194 &2.108 &1.456 &1.424 &1.867 &1.015 &1.569\\
\midrule
MLP  &2.524 &2.214 &1.899 &1.680 &1.469 &1.597 &0.937 &0.981 &1.060 &1.441 &0.802 &1.159\\
LSTM  &2.290 &2.210 &1.750 &1.680 &1.346 &1.304 &0.724 &0.779 &1.219 &1.296 &0.762 &0.558\\
MMIM  &2.290 &2.092 &1.779 &1.598 &1.287 &1.133 &0.718 &0.622 &\underline{0.975} &1.081 &0.500& 0.510\\
MDRM  &\underline{2.065} &2.511 &1.748 &1.597 &1.281 &1.578 &0.683 &0.612 &1.183 &1.627 &0.769 &0.512\\
HTML  &2.296 &2.133 &1.771 &1.611 &1.302 &1.127 &0.766 &0.609 &0.988 &1.118 &0.588 &0.498\\
MULT  &2.073 &2.179 &1.768 &1.605 &1.288 &1.133 &\underline{0.672} &0.742 &1.022 &1.018 &0.549& 0.497\\
MPCNet & 2.233& \underline{2.089} & \underline{1.732} & \underline{1.594} & \underline{1.269} &\underline{1.046} & 0.806 & \underline{0.607} & 1.176 & \underline{1.001} & \underline{0.469} & \underline{0.470} \\
\midrule
\textbf{MANAGER}&$\textbf{1.819}^{\star}$ &$\textbf{1.725}^{\star}$&$\textbf{1.608}^{\star}$ & $\textbf{1.471}^{\star}$ &$\textbf{1.126}^{\star}$ & $\textbf{0.813}^{\star}$ & $\textbf{0.584}^{\star}$ & $\textbf{0.572}^{\star}$ &$\textbf{0.906}^{\star}$ &$\textbf{0.957}^{\star}$ &$\textbf{0.416}^{\star}$ & $\textbf{0.402}^{\star}$ \\
\bottomrule
\end{tabular}}
\subcaption{ Stock Indices and Currency Exchange Rate
}
\end{subtable}

\begin{subtable}[t]{0.5\linewidth}
\resizebox{\linewidth}{!}{
\begin{tabular}{l|cccc|cccc|cccc}
\toprule
\multirow{2}{*}{\textbf{Model}} & \multicolumn{4}{c|}{\textbf{Gold}}                                                                         & \multicolumn{4}{c|}{\textbf{10-Year Bond Yield}} & \multicolumn{4}{c}{\textbf{3-Month Bond Yield}}  \\ 
\cmidrule{2-13} 
& $\mathbf{MSE_1}$ &$\mathbf{MSE_3}$ & $\mathbf{MSE_7}$ & $\mathbf{MSE_{15}}$ & $\mathbf{MSE_1}$ & $\mathbf{MSE_3}$ & $\mathbf{MSE_7}$ &$\mathbf{MSE_{15}}$ & $\mathbf{MSE_1}$ & $\mathbf{MSE_3}$ & $\mathbf{MSE_7}$ &$\mathbf{MSE_{15}}$ \\
\midrule
HistPrice  &3.193 &3.039 &2.675 &2.683 &4.132 &4.020 &3.472 &3.334 &3.899 &3.665 &3.063 &2.913\\
P-SVM  &2.568 &2.543 &1.967 &2.104 &3.212 &3.589 &2.986 &3.141 &3.235 &3.143 &2.922 &2.874\\
P-LSTM  &1.965 &1.998 &1.043 &1.764 &2.212 &1.699 &2.340 &1.453 &3.433 &2.909 &2.678 &2.477\\
\midrule
MLP  &1.431 &1.654 &0.904 &0.955 &1.811 &1.743 &1.288 &1.382 &2.582 &2.523 &2.239 &2.231\\
LSTM &1.472 &1.484 &0.703& 0.508 &1.735 &1.801 &1.169 &1.235 &2.421 &2.439 &2.044 &2.013\\
MMIM  &1.292 &1.292 &0.565 &0.486 &1.698 &1.604 &1.080 &1.053 &2.345 &2.392 &1.977 &1.902\\
MDRM  &1.436 &1.843 &0.710 &0.483 &1.729 &1.699 &1.126 &1.223 &2.406 &2.622 &2.096 &1.993\\
HTML  &\underline{1.277} &1.291 &0.589 &0.524 &\underline{1.685} &1.612 &1.103 &1.149 &2.342 &2.356 &1.962 &1.998\\
MULT  &1.314 &1.335 &0.579 &0.503 &2.122 &1.837 &1.104 &\underline{1.037} &\underline{1.174} &2.515 &1.973 &1.903\\
MPCNet & 1.342 & \underline{1.275} &\underline{0.562}  &\underline{0.477}  &1.767  &\underline{1.602}  &\underline{0.979}  &1.142  &2.431  &\underline{2.319}  &\underline{1.948}  &\underline{1.901}  \\
\midrule
\textbf{MANAGER} &$\textbf{1.106}^{\star}$ &$\textbf{1.144}^{\star}$ &$\textbf{0.527}^{\star}$ & $\textbf{0.419}^{\star}$  &$\textbf{1.452}^{\star}$  &$\textbf{1.574}^{\star}$  &$\textbf{0.825}^{\star}$  &$\textbf{0.917}^{\star}$ &$\textbf{1.049}^{\star}$ &$\textbf{2.076}^{\star}$ &$\textbf{1.804}^{\star}$ &$\textbf{1.276}^{\star}$ \\
\bottomrule
\end{tabular}}
\subcaption{Gold Prices, Long-term (10-Years) and Short-term (3-Months) Bonds}
\end{subtable}

\end{floatrow}

\end{table*}

\begin{table*}
\centering
\caption{Ablation study results of our proposed MANAGER for movement prediction. The best results are highlighted in boldface.}
\label{Tab:Ablation Study movement}
\begin{floatrow}
\begin{subtable}[t]{0.5\linewidth}
\resizebox{\linewidth}{!}{
\begin{tabular}{l|cccc|cccc|cccc}
\toprule
\multirow{2}{*}{\textbf{Model}} & \multicolumn{4}{c|}{\textbf{Stock Index (Small)}}                                                                         & \multicolumn{4}{c|}{\textbf{Stock Index (Large)}} & \multicolumn{4}{c}{\textbf{Currency Exchange Rate}}  \\ 
\cmidrule{2-13} 
& \textbf{F1\_1} & \textbf{F1\_3} & \textbf{F1\_7} & \textbf{F1\_15} & \textbf{F1\_1} & \textbf{F1\_3} & \textbf{F1\_7} & \textbf{F1\_15} & \textbf{F1\_1} & \textbf{F1\_3} & \textbf{F1\_7} & \textbf{F1\_15} \\
\midrule
w/o-Text &0.479 &0.590 &0.513 &0.544  &0.437  & 0.556& 0.436 &0.549 &0.483  &0.510  & 0.433 &0.609\\
w/o-Knowledge &0.530 &0.677 &0.592 &0.631  &0.485  &0.630 &0.574  &0.628 &0.547  &0.591  &0.497  &0.670\\
w/o-Video &0.509  &0.664 &0.581  &0.627  &0.477  &0.606 &0.562 &0.611 &0.530  &0.573  &0.484 &0.657  \\
w/o-Audio &0.527 &0.671 &0.586 &0.612 &0.493  &0.639 &0.570 &0.620 &0.546 &0.588  &0.494 &0.660\\
w/o-Graph  &0.533 &0.679 &0.491 &0.608 &0.492 &0.611 &0.560 &0.627 &0.532  &0.570 &0.483 &0.654 \\
w/-FullGraph  &0.499 &0.582 &0.548 &0.601 &0.429 &0.627 &0.451 &0.595   &0.510  &0.521 &0.448 &0.639\\
\midrule
\textbf{MANAGER}& \textbf{0.548} & \textbf{0.694} & \textbf{0.610} & \textbf{0.659} &\textbf{0.517} & \textbf{0.652} & \textbf{0.589} & \textbf{0.646} & \textbf{0.564} & \textbf{0.608} & \textbf{0.511} & \textbf{0.681} \\
\bottomrule
\end{tabular}}
\subcaption{Stock Indices and Currency Exchange Rate}
\end{subtable}

\begin{subtable}[t]{0.5\linewidth}
\resizebox{\linewidth}{!}{
\begin{tabular}{l|cccc|cccc|cccc}
\toprule
\multirow{2}{*}{\textbf{Model}} & \multicolumn{4}{c|}{\textbf{Gold}}                                                                         & \multicolumn{4}{c|}{\textbf{10-Year Bond Yield}} & \multicolumn{4}{c}{\textbf{3-Month Bond Yield}}  \\ 
\cmidrule{2-13} 
& \textbf{F1\_1} & \textbf{F1\_3} & \textbf{F1\_7} & \textbf{F1\_15} & \textbf{F1\_1} & \textbf{F1\_3} & \textbf{F1\_7} & \textbf{F1\_15} & \textbf{F1\_1} & \textbf{F1\_3} & \textbf{F1\_7} & \textbf{F1\_15} \\
\midrule
w/o-Text &0.369 &0.571 &0.448 &0.525  &0.498  &0.328 &0.467  &0.583 &0.460  &0.516  &0.428  &0.502\\
w/o-Knowledge &0.470 &0.659 &0.488 &0.656  &0.593  &0.370 &0.562  &0.627 &0.504  &0.576  &0.498  &0.557\\
w/o-Video &0.451  &0.639 &0.471  &0.646  &0.589  &0.352 &0.550 &0.617 &0.501  &0.537  &0.500 &0.545  \\
w/o-Audio &0.456 &0.662 &0.479 &0.631 & 0.573&0.362 &0.554 &0.619 &0.496 &0.553  &0.481 &0.521\\
w/o-Graph &0.457 &0.681 &0.457 &0.650  &0.593 &0.359 &0.559 &0.624  &0.503  &0.551 &0.487 &0.540\\
w/-FullGraph  &0.397 &0.592 &0.461 &0.620 &0.519 &0.317 &0.512 &0.607 &0.495  &0.532   &0.413 &0.518\\
\midrule
\textbf{MANAGER} &\textbf{0.486} &\textbf{0.696} &\textbf{0.507} &\textbf{0.672}  &\textbf{0.612}  &\textbf{0.391} &\textbf{0.587}  &\textbf{0.649} &\textbf{0.521} &\textbf{0.583} &\textbf{0.519} &\textbf{0.574} \\
\bottomrule
\end{tabular}}
\subcaption{Gold Prices, Long-term (10-Years) and Short-term (3-
Months) Bonds}
\end{subtable}

\end{floatrow}
\end{table*}

\begin{table*}
\centering
\caption{Ablation study results of our proposed MANAGER for volatility prediction. The best results are highlighted in boldface.}
\label{Tab:Ablation Study volatility}
\begin{floatrow}
\begin{subtable}[t]{0.5\linewidth}
\resizebox{\linewidth}{!}{
\begin{tabular}{l|cccc|cccc|cccc}
\toprule
\multirow{2}{*}{\textbf{Model}} & \multicolumn{4}{c|}{\textbf{Stock Index (Small)}}                                                                         & \multicolumn{4}{c|}{\textbf{Stock Index (Large)}} & \multicolumn{4}{c}{\textbf{Currency Exchange Rate}}  \\ 
\cmidrule{2-13} 
& $\mathbf{MSE_1}$ &$\mathbf{MSE_3}$ & $\mathbf{MSE_7}$ & $\mathbf{MSE_{15}}$ & $\mathbf{MSE_1}$ & $\mathbf{MSE_3}$ & $\mathbf{MSE_7}$ &$\mathbf{MSE_{15}}$ & $\mathbf{MSE_1}$ & $\mathbf{MSE_3}$ & $\mathbf{MSE_7}$ & $\mathbf{MSE_{15}}$ \\
\midrule
w/o-Text &2.146 &2.048 &1.795 &1.613  &1.379  & 1.016 & 0.879 &0.926 &1.198  &1.348  & 0.702 &0.617\\
w/o-Knowledge &1.930 &1.908 &1.784 &1.536  &1.291  &0.905 &0.670  &0.713 &0.977  &1.019  &0.474  &0.566\\
w/o-Video &2.101  &1.997 &1.891  &1.741  &1.324  &1.089 &0.851 &0.854 &1.122  &1.141  &0.668 &0.609  \\
w/o-Audio &1.959 &1.893 &1.748 &1.609 &1.237  &1.003 &0.776 &0.790 &1.035 &1.124  &0.603 &0.593\\
w/o-Graph  &1.941 &1.929 &1.754 &1.712 &1.331 &1.004 &0.750 &0.755 &1.136  &1.180 &0.603 &0.449 \\
w/-FullGraph  &2.144 &1.962 &1.919 &1.791 &1.495 &1.022 &0.898 &0.904   &1.175  &1.306 &0.681 &0.640\\
\midrule
\textbf{MANAGER}&\textbf{1.819} &\textbf{1.725}&\textbf{1.608} & \textbf{1.471} &\textbf{1.126} & \textbf{0.813} & \textbf{0.584} & \textbf{0.572} &\textbf{0.906} &\textbf{0.957} &\textbf{0.416} & \textbf{0.402} \\
\bottomrule
\end{tabular}}
\subcaption{Stock Indices and Currency Exchange Rate}
\end{subtable}

\begin{subtable}[t]{0.5\linewidth}
\resizebox{\linewidth}{!}{
\begin{tabular}{l|cccc|cccc|cccc}
\toprule
\multirow{2}{*}{\textbf{Model}} & \multicolumn{4}{c|}{\textbf{Gold}}                                                                         & \multicolumn{4}{c|}{\textbf{10-Year Bond Yield}} & \multicolumn{4}{c}{\textbf{3-Month Bond Yield}}  \\ 
\cmidrule{2-13} 
& $\mathbf{MSE_1}$ &$\mathbf{MSE_3}$ & $\mathbf{MSE_7}$ & $\mathbf{MSE_{15}}$ & $\mathbf{MSE_1}$ & $\mathbf{MSE_3}$ & $\mathbf{MSE_7}$ &$\mathbf{MSE_{15}}$ & $\mathbf{MSE_1}$ & $\mathbf{MSE_3}$ & $\mathbf{MSE_7}$ & $\mathbf{MSE_{15}}$ \\
\midrule
w/o-Text &1.451 &1.400 &0.796 &0.794  &1.812  &1.910 &1.110  &1.173 &1.400  &2.403  &2.090  &1.498\\
w/o-Knowledge &1.222 &1.215 &0.593 &0.427  &1.594  &1.708 &0.886  &1.092 &1.237  &2.211  &1.900  &1.467\\
w/o-Video &2.184  &1.351 &0.576  &0.704  &1.641  &1.850 &1.070 &1.101 &1.344  &2.280  &1.979 &1.510  \\
w/o-Audio &1.307 &1.358 &0.639 &0.549 &1.588 &1.758 &1.024 &1.127 &1.264 &2.213  &1.943 &1.419\\
w/o-Graph &1.347 &1.278 &0.742 &0.540  &0.674 &1.861 &1.013 &1.164  &1.275  &2.226 &1.937 &1.494\\
w/-FullGraph  &1.365 &1.367 &0.795 &0.680 &1.816 &1.834 &1.073 &1.168 &1.234  &2.373   &2.075 &1.463\\
\midrule
\textbf{MANAGER} &\textbf{1.106} &\textbf{1.144} &\textbf{0.527} &\textbf{0.419}  &\textbf{1.452}  &\textbf{1.574}  &\textbf{0.825}  &\textbf{0.917} &\textbf{1.049} &\textbf{2.076} &\textbf{1.804} &\textbf{1.276} \\
\bottomrule
\end{tabular}}
\subcaption{Gold Prices, Long-term (10-Years) and Short-term (3-
Months) Bonds}
\end{subtable}

\end{floatrow}
\end{table*}

%% file: Chapter/4_experiment.tex
\vspace{-0.4cm}
\section{Experiment}
\subsection{Dataset}
In this work, we conducted extensive experiments on Monopoly~\cite{Mathur2022MONOPOLYFP} dataset for financial prediction. It is a collection of public monetary conference call videos along with their corresponding audio recordings and text transcripts released by six international banks between $2009$ and $2022$. Overall, it consists of $24,180$ samples, and each sample includes the corresponding text, video and audio clips with the annotated price movement and volatility. We adopted the original dataset split setting, the ratio of data split for training/validation/testing sets is $7:1:2$. 
\subsection{Experimental Setup}
We adopted ChatGLM2\footnote{\url{https://huggingface.co/THUDM/chatglm2-6b}.} as the backbone of our model. 
The total number of tokens in each sample, i.e., $N_t$ is unified to $768$. The feature dimension $D$ is set to $768$.
We used AdamW~\cite{DBLP:journals/corr/abs-1711-05101} as the optimizer and set the learning rate of GCN layers to 1e-3. Following~\citeauthor{Mathur2022MONOPOLYFP}~(\citeyear{Mathur2022MONOPOLYFP}), we use a learning rate of 1e-4 for movement prediction and 1e-3 for volatility prediction, respectively.
The batch size is set to $1$ due to GPU limitation, and the maximum number of epochs for model training is set to $10$.
Following the previous work, we employed mean squared error (MSE) to evaluate the predicted volatility and used F1 score to measure the predicted price movement, respectively, for $\tau \in \{1,3,7,15\}$.

\subsection{Baseline methods}
\subsubsection{Text-only baselines}
\begin{itemize}
    \item \textbf{HistPrice}~\cite{Du2007DoesPV}. It utilizes the ARIMA model to perform
regression/classification.
    \item \textbf{P-SVM}~\cite{Chatzis2018ForecastingSM}. This model applies Support Vector Regression (SVR) and
Classifiers (SVC) for volatility and price movement prediction, respectively.
    \item \textbf{P-LSTM}~\cite{Yu2018ForecastingSP}. It uses LSTM to extract forecast patterns from 30-day historical price time-series.
\end{itemize}
\subsubsection{Multimodal baselines}
\begin{itemize}
    \item \textbf{MLP}~\cite{Tolstikhin2021MLPMixerAA}. It is a simple multi-layer perceptron where multimodal features are aggregated across a time series and concatenated for prediction.
    \item \textbf{LSTM}~\cite{Poria2017ContextDependentSA}. It feeds the multimodal time series to individual LSTMs and averages them before the final prediction.
    \item \textbf{MMIM}~\cite{Han2021ImprovingMF}. In this model, LSTMs are employed to encode the video and audio sequences, while BERT is utilized for text processing. Subsequently, the encoded features are fused for prediction.
    \item \textbf{MDRM}~\cite{Qin2019WhatYS}. It adopts BiLSTM layers to encode unimodal sequences, which are then fused together
    for prediction.
    \item \textbf{HTML}~\cite{Yang2020HTMLHT}. HTML 
    utilizes fused multimodal feature representations before passing through Transformer layers for final prediction.
    \item \textbf{MULT}~\cite{Tsai2019MultimodalTF}. It employs transformer encoders to align text, video, and audio sequences 
    for prediction.
    \item \textbf{MPCNet}~\cite{Mathur2022MONOPOLYFP}. It adopts cross-modal transformer blocks and modality-specific attention fusion for prediction.
\end{itemize}
\subsection{Experimental results}
We reported the experiment results in Table~\ref{Tab:Comparison movement} and Table~\ref{Tab:Comparison volatility}. From the above tables, we have the following observations. 1) Our model \textbf{MANAGER} consistently exceeds all the baselines in terms of all the metrics for both price movement and volatility prediction, which thoroughly demonstrates the superiority of our model. 
2) Overall, the second best model is always multimodal baseline 
which verifies that the video and audio modalities can provide useful information for the financial prediction. 
3) Notably, multimodal models not always outperform text-only models. For example, \textbf{HistPrice}, \textbf{P-SVM} and \textbf{P-LSTM} exceed \textbf{MLP} in the movement prediction of Stock Index (Large). It imples that improper use of non-verbal information in video and audio may lead to worse performance.  

\section{Analyses}
\subsection{Ablation Study}
We introduced the following variants to explore the contribution of each component. 
\begin{itemize}
\item\textbf{w/o-Text}, \textbf{w/o-Knowledge}, \textbf{w/o-Video} \textbf{w/o-Audio} and \textbf{w/o-Graph}.
To prove the effectiveness of the input text, inferred knowledge, video, audio and constructed knowledge-enhanced cross-modal graph, we eliminated  the text, external financial knowledge, video, audio and graph in these variants, respectively.
\item \textbf{w/-FullGraph}. To further investigate the semantic relations of our knowledge-enhanced cross-modal graph, we erased all the semantic relations and transformed the semantic graph to a full connection graph.
\end{itemize}
The ablation study results are shown in Table~\ref{Tab:Ablation Study movement} and Table~\ref{Tab:Ablation Study volatility}. From this table, we have the following observations. 
1) w/o-Text performs terribly compared with MANAGER. This is reasonable since the caption is the main source for delivering information to predict the price movement or volatility.
2) MANAGER exceeds w/o-Knowledge. It proves that external knowledge in the financial domain can assist in comprehending the context. 
3) MANAGER consistently outperforms w/o-Video and w/o-audio across different evaluation metrics. It demonstrates the non-verbal information residing in the video and audio can improve context comprehension and hence boost financial prediction. 
4)w/o-Text performs worse than both w/o-Video and w/o-Audio. It implies that the given text contributes more to the financial prediction than video and audio.
5) MANAGER outperforms w/o-Graph, denoting that the graphs are essential to capture the semantic relations
among text, knowledge, video and audio, which help comprehend the cross-modal context.
And 6) w/-FullGraph performs worse than MANAGER, which verifies the utility of proposed semantic relations.
\begin{figure}[h]
    \centering
    \includegraphics[scale=0.44]{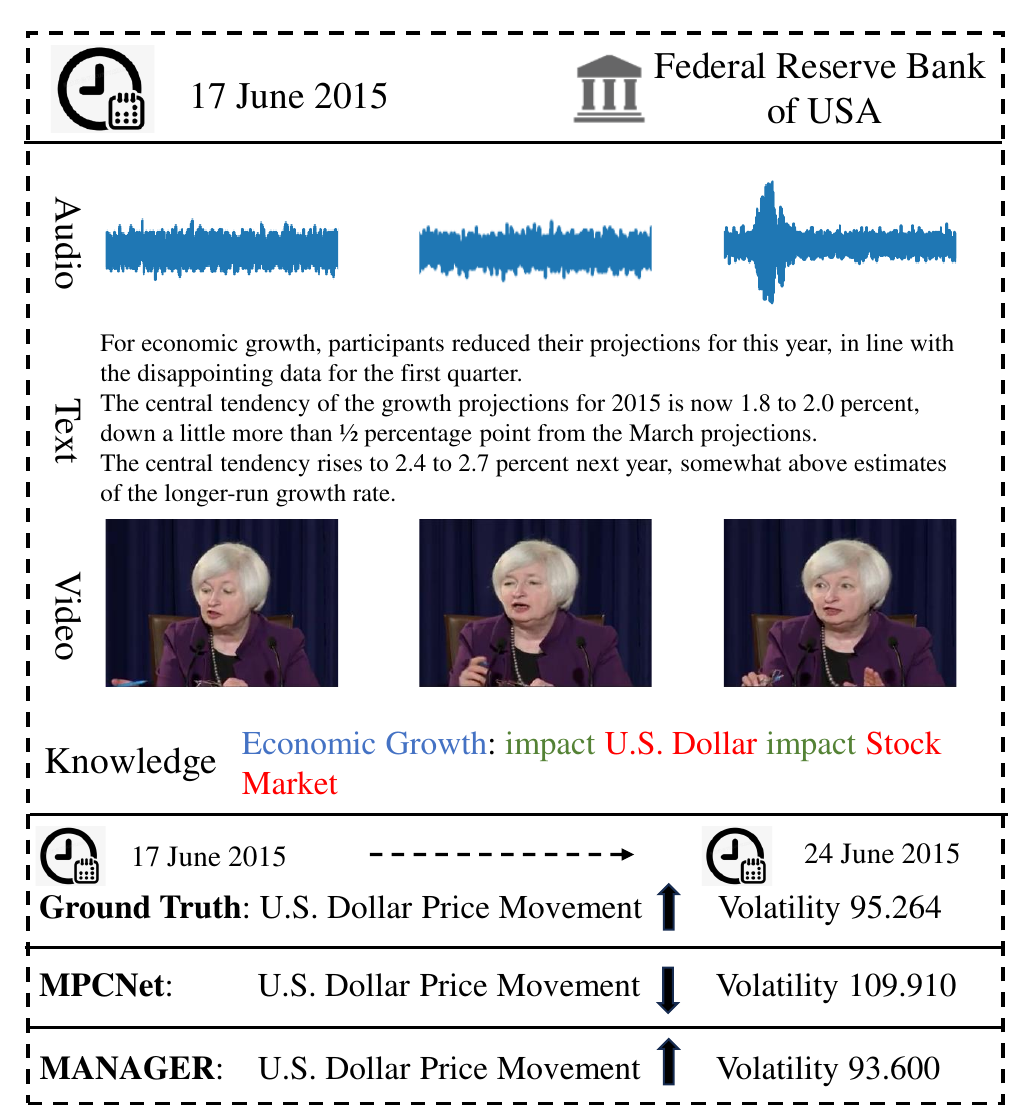}
     \vspace{-0.1cm}
    \caption{Comparison between the results predicted by MANAAGER and the best baseline MPCNet on one testing sample.}
    \label{fig:case}
\end{figure}
\vspace{-0.4cm}
\vspace{0.12cm}
\subsection{Case Study}
To get an intuitive understanding of how our model works on financial prediction from MPC calls, we showed one testing sample in Figure~\ref{fig:case} due to the limited space. For comparison, we also displayed the prediction results of the best baseline MPCNet.

As you can see, our MANAGER predicted the price movement of U.S. Dollar correctly, while MPCNet failed. In addition, the volatility $93.600$ forecasted by MANAGER is closer to the ground truth $95.264$ than $109.910$ predicted by MPCNet. This may be attributed to the fact that the external knowledge (e.g., relation: ``impact'', entity: ``U.S. Dollar'' and ``Stock Market'') inferred by the entity ``Economic Growth'' may guide our model to pay attention to ``Economic Growth'' existed in the text, since it may provide some useful information for the price movement or volatility of U.S. Dollar. Overall, this case shows the benefit of incorporating external knowledge into the context of financial prediction from MPC calls. 

%% file: Chapter/5_conclusion.tex
\section{Conclusion and Future Work}
In this work, we propose a novel modal-adaptive knowledge-enhanced graph-based financial prediction scheme. Experimental results on a
public dataset demonstrate the superiority of our
model over existing cutting-edge methods, and validate the advantage of utilizing external knowledge in the financial domain, as well as the benefit of constructing the knowledge-enhanced cross-modal graph to characterize the intra-modal and inter-modal relations among the multiple input (i.e., text, external knowledge, video and audio). In the future, we plan to explore the Multimodal Large Language Models, such as VisualGLM, in financial prediction.

%% file: main.bbl
\begin{thebibliography}{39}
\expandafter\ifx\csname natexlab\endcsname\relax\def\natexlab#1{#1}\fi

\bibitem[{Boukus and Rosenberg(2006)}]{Boukus2006TheIC}
Ellyn~R. Boukus and Joshua~V. Rosenberg. 2006.
\newblock The information content of fomc minutes.
\newblock \emph{Monetary Economics}.

\bibitem[{Brown et~al.(2020)Brown, Mann, Ryder, Subbiah, Kaplan, Dhariwal,
  Neelakantan, Shyam, Sastry, Askell, Agarwal, Herbert-Voss, Krueger, Henighan,
  Child, Ramesh, Ziegler, Wu, Winter, Hesse, Chen, Sigler, Litwin, Gray, Chess,
  Clark, Berner, McCandlish, Radford, Sutskever, and
  Amodei}]{Brown2020LanguageMA}
Tom~B. Brown, Benjamin Mann, Nick Ryder, Melanie Subbiah, Jared Kaplan,
  Prafulla Dhariwal, Arvind Neelakantan, Pranav Shyam, Girish Sastry, Amanda
  Askell, Sandhini Agarwal, Ariel Herbert-Voss, Gretchen Krueger, T.~J.
  Henighan, Rewon Child, Aditya Ramesh, Daniel~M. Ziegler, Jeff Wu, Clemens
  Winter, Christopher Hesse, Mark Chen, Eric Sigler, Mateusz Litwin, Scott
  Gray, Benjamin Chess, Jack Clark, Christopher Berner, Sam McCandlish, Alec
  Radford, Ilya Sutskever, and Dario Amodei. 2020.
\newblock Language models are few-shot learners.
\newblock \emph{ArXiv}.

\bibitem[{Cai et~al.(2021)Cai, Camara, and Capel}]{Cai2021ItsNA}
Yongbin Cai, Santiago Camara, and Nicholas Capel. 2021.
\newblock It's not always about the money, sometimes it's about sending a
  message: Evidence of informational content in monetary policy announcements.

\bibitem[{Cao(2021)}]{Cao2021AIIF}
Longbing Cao. 2021.
\newblock Ai in finance: Challenges, techniques, and opportunities.
\newblock \emph{ACM Computing Surveys (CSUR)}.

\bibitem[{Chatzis et~al.(2018)Chatzis, Siakoulis, Petropoulos, Stavroulakis,
  and Vlachogiannakis}]{Chatzis2018ForecastingSM}
Sotirios~P. Chatzis, Vasilis Siakoulis, Anastasios Petropoulos, Evangelos
  Stavroulakis, and Nikos~E. Vlachogiannakis. 2018.
\newblock Forecasting stock market crisis events using deep and statistical
  machine learning techniques.
\newblock \emph{Expert Syst. Appl.}

\bibitem[{Devlin et~al.(2019)Devlin, Chang, Lee, and
  Toutanova}]{DBLP:conf/naacl/DevlinCLT19}
Jacob Devlin, Ming{-}Wei Chang, Kenton Lee, and Kristina Toutanova. 2019.
\newblock {BERT:} pre-training of deep bidirectional transformers for language
  understanding.
\newblock In \emph{Proceedings of the Conference of the North American Chapter
  of the Association for Computational Linguistics: Human Language
  Technologies}, pages 4171--4186. ACL.

\bibitem[{Du and Budescu(2007)}]{Du2007DoesPV}
Ning Du and David~V. Budescu. 2007.
\newblock Does past volatility affect investors' price forecasts and confidence
  judgements?
\newblock \emph{International Journal of Forecasting}.

\bibitem[{Du et~al.(2022)Du, Qian, Liu, Ding, Qiu, Yang, and Tang}]{du2022glm}
Zhengxiao Du, Yujie Qian, Xiao Liu, Ming Ding, Jiezhong Qiu, Zhilin Yang, and
  Jie Tang. 2022.
\newblock Glm: General language model pretraining with autoregressive blank
  infilling.
\newblock In \emph{Proceedings of the 60th Annual Meeting of the Association
  for Computational Linguistics}, pages 320--335.

\bibitem[{Han et~al.(2021)Han, Chen, and Poria}]{Han2021ImprovingMF}
Wei Han, Hui Chen, and Soujanya Poria. 2021.
\newblock Improving multimodal fusion with hierarchical mutual information
  maximization for multimodal sentiment analysis.
\newblock \emph{ArXiv}.

\bibitem[{Hsu et~al.(2021)Hsu, Bolte, Tsai, Lakhotia, Salakhutdinov, and rahman
  Mohamed}]{Hsu2021HuBERTSS}
Wei-Ning Hsu, Benjamin Bolte, Yao-Hung~Hubert Tsai, Kushal Lakhotia, Ruslan
  Salakhutdinov, and Abdel rahman Mohamed. 2021.
\newblock Hubert: Self-supervised speech representation learning by masked
  prediction of hidden units.
\newblock \emph{IEEE/ACM Transactions on Audio, Speech, and Language
  Processing}.

\bibitem[{Hu et~al.(2021)Hu, Shen, Wallis, Allen-Zhu, Li, Wang, and
  Chen}]{Hu2021LoRALA}
J.~Edward Hu, Yelong Shen, Phillip Wallis, Zeyuan Allen-Zhu, Yuanzhi Li, Shean
  Wang, and Weizhu Chen. 2021.
\newblock Lora: Low-rank adaptation of large language models.
\newblock \emph{ArXiv}.

\bibitem[{Jiang(2020)}]{Jiang2020ApplicationsOD}
Weiwei Jiang. 2020.
\newblock Applications of deep learning in stock market prediction: recent
  progress.
\newblock \emph{Expert Syst. Appl.}

\bibitem[{Jing et~al.(2023)Jing, Song, Ouyang, Jia, and
  Nie}]{Jing2023MultisourceSG}
Liqiang Jing, Xuemeng Song, Kun Ouyang, Mengzhao Jia, and Liqiang Nie. 2023.
\newblock Multi-source semantic graph-based multimodal sarcasm explanation
  generation.
\newblock In \emph{Annual Meeting of the Association for Computational
  Linguistics}. ACL.

\bibitem[{Kipf and Welling(2017)}]{DBLP:conf/iclr/KipfW17}
Thomas~N. Kipf and Max Welling. 2017.
\newblock Semi-supervised classification with graph convolutional networks.
\newblock In \emph{International Conference on Learning Representations}.
  OpenReview.net.

\bibitem[{Lewellen(2003)}]{Lewellen2003FinancingDW}
Katharina Lewellen. 2003.
\newblock Financing decisions when managers are risk averse.
\newblock \emph{MIT Sloan School of Management Working Paper Series}.

\bibitem[{Li(2023)}]{1}
Xiaohui~Victor Li. 2023.
\newblock Findkg: Dynamic knowledge graph with large language models for global
  finance.
\newblock pages 1--64. SSRN.

\bibitem[{Liu and Chilton(2021)}]{Liu2021DesignGF}
Vivian Liu and Lydia~B. Chilton. 2021.
\newblock Design guidelines for prompt engineering text-to-image generative
  models.
\newblock \emph{Proceedings of the 2022 CHI Conference on Human Factors in
  Computing Systems}.

\bibitem[{Liu et~al.(2020)Liu, Huang, Huang, Li, and Zhao}]{ijcai2020p622}
Zhuang Liu, Degen Huang, Kaiyu Huang, Zhuang Li, and Jun Zhao. 2020.
\newblock Finbert: A pre-trained financial language representation model for
  financial text mining.
\newblock In \emph{Proceedings of the Twenty-Ninth International Joint
  Conference on Artificial Intelligence, {IJCAI-20}}, pages 4513--4519.
  International Joint Conferences on Artificial Intelligence Organization.

\bibitem[{Loshchilov and Hutter(2017)}]{DBLP:journals/corr/abs-1711-05101}
Ilya Loshchilov and Frank Hutter. 2017.
\newblock Fixing weight decay regularization in adam.
\newblock \emph{CoRR}, abs/1711.05101.

\bibitem[{Mathur et~al.(2022)Mathur, Neerkaje, Chhibber, Sawhney, Guo,
  Dernoncourt, Dutta, and Manocha}]{Mathur2022MONOPOLYFP}
Puneet Mathur, Atula~Tejaswi Neerkaje, Malika Chhibber, Ramit Sawhney, Fuming
  Guo, Franck Dernoncourt, Sanghamitra Dutta, and Dinesh Manocha. 2022.
\newblock Monopoly: Financial prediction from monetary policy conference videos
  using multimodal cues.
\newblock \emph{Proceedings of the 30th ACM International Conference on
  Multimedia}.

\bibitem[{Ouyang et~al.(2024)Ouyang, Jing, Song, Liu, Hu, and
  Nie}]{DBLP:journals/corr/abs-2402-03658}
Kun Ouyang, Liqiang Jing, Xuemeng Song, Meng Liu, Yupeng Hu, and Liqiang Nie.
  2024.
\newblock Sentiment-enhanced graph-based sarcasm explanation in dialogue.
\newblock \emph{CoRR}, abs/2402.03658.

\bibitem[{Ouyang et~al.(2022)Ouyang, Wu, Jiang, Almeida, Wainwright, Mishkin,
  Zhang, Agarwal, Slama, Ray, Schulman, Hilton, Kelton, Miller, Simens, Askell,
  Welinder, Christiano, Leike, and Lowe}]{Ouyang2022TrainingLM}
Long Ouyang, Jeff Wu, Xu~Jiang, Diogo Almeida, Carroll~L. Wainwright, Pamela
  Mishkin, Chong Zhang, Sandhini Agarwal, Katarina Slama, Alex Ray, John
  Schulman, Jacob Hilton, Fraser Kelton, Luke~E. Miller, Maddie Simens, Amanda
  Askell, Peter Welinder, Paul~Francis Christiano, Jan Leike, and Ryan~J. Lowe.
  2022.
\newblock Training language models to follow instructions with human feedback.
\newblock \emph{ArXiv}.

\bibitem[{Poria et~al.(2017)Poria, Cambria, Hazarika, Majumder, Zadeh, and
  Morency}]{Poria2017ContextDependentSA}
Soujanya Poria, E.~Cambria, Devamanyu Hazarika, Navonil Majumder, Amir Zadeh,
  and Louis-Philippe Morency. 2017.
\newblock Context-dependent sentiment analysis in user-generated videos.
\newblock In \emph{Annual Meeting of the Association for Computational
  Linguistics}.

\bibitem[{Qin and Yang(2019)}]{Qin2019WhatYS}
Yu~Qin and Yi~Yang. 2019.
\newblock What you say and how you say it matters: Predicting stock volatility
  using verbal and vocal cues.
\newblock In \emph{Annual Meeting of the Association for Computational
  Linguistics}.

\bibitem[{Sawhney et~al.(2021)Sawhney, Goyal, Goel, Mathur, and
  Shah}]{Sawhney2021MultimodalMM}
Ramit Sawhney, Mihir Goyal, Prakhar Goel, Puneet Mathur, and Rajiv~Ratn Shah.
  2021.
\newblock Multimodal multi-speaker merger \& acquisition financial modeling: A
  new task, dataset, and neural baselines.
\newblock In \emph{Annual Meeting of the Association for Computational
  Linguistics}.

\bibitem[{Sawhney et~al.(2020)Sawhney, Mathur, Mangal, Khanna, Shah, and
  Zimmermann}]{Sawhney2020MultimodalMF}
Ramit Sawhney, Puneet Mathur, Ayush Mangal, Piyush Khanna, Rajiv~Ratn Shah, and
  Roger Zimmermann. 2020.
\newblock Multimodal multi-task financial risk forecasting.
\newblock \emph{Proceedings of the 28th ACM International Conference on
  Multimedia}.

\bibitem[{Shapiro and Wilson(2019)}]{Shapiro2019TakingTF}
Adam~Hale Shapiro and Daniel~J. Wilson. 2019.
\newblock Taking the fed at its word: A new approach to estimating central bank
  objectives using text analysis.
\newblock \emph{Federal Reserve Bank of San Francisco, Working Paper Series}.

\bibitem[{Tolstikhin et~al.(2021)Tolstikhin, Houlsby, Kolesnikov, Beyer, Zhai,
  Unterthiner, Yung, Keysers, Uszkoreit, Lucic, and
  Dosovitskiy}]{Tolstikhin2021MLPMixerAA}
Ilya~O. Tolstikhin, Neil Houlsby, Alexander Kolesnikov, Lucas Beyer, Xiaohua
  Zhai, Thomas Unterthiner, Jessica Yung, Daniel Keysers, Jakob Uszkoreit,
  Mario Lucic, and Alexey Dosovitskiy. 2021.
\newblock Mlp-mixer: An all-mlp architecture for vision.
\newblock In \emph{Neural Information Processing Systems}.

\bibitem[{Tsai et~al.(2019)Tsai, Bai, Liang, Kolter, Morency, and
  Salakhutdinov}]{Tsai2019MultimodalTF}
Yao-Hung~Hubert Tsai, Shaojie Bai, Paul~Pu Liang, J.~Zico Kolter,
  Louis-Philippe Morency, and Ruslan Salakhutdinov. 2019.
\newblock Multimodal transformer for unaligned multimodal language sequences.
\newblock \emph{Proceedings of the conference. Association for Computational
  Linguistics. Meeting}.

\bibitem[{Wang et~al.(2023{\natexlab{a}})Wang, Yang, and
  Wang}]{Wang2023FinGPTIT}
Neng Wang, Hongyang Yang, and Chris Wang. 2023{\natexlab{a}}.
\newblock Fingpt: Instruction tuning benchmark for open-source large language
  models in financial datasets.
\newblock \emph{ArXiv}.

\bibitem[{Wang et~al.(2023{\natexlab{b}})Wang, Yang, and
  Wang}]{wang2023fingptbenchmark}
Neng Wang, Hongyang Yang, and Christina~Dan Wang. 2023{\natexlab{b}}.
\newblock Fingpt: Instruction tuning benchmark for open-source large language
  models in financial datasets.
\newblock \emph{NeurIPS Workshop on Instruction Tuning and Instruction
  Following}.

\bibitem[{Wang et~al.(2022)Wang, Bao, Dong, Bjorck, Peng, Liu, Aggarwal,
  Mohammed, Singhal, Som, and Wei}]{DBLP:journals/corr/abs-2208-10442}
Wenhui Wang, Hangbo Bao, Li~Dong, Johan Bjorck, Zhiliang Peng, Qiang Liu, Kriti
  Aggarwal, Owais~Khan Mohammed, Saksham Singhal, Subhojit Som, and Furu Wei.
  2022.
\newblock Image as a foreign language: Beit pretraining for all vision and
  vision-language tasks.
\newblock \emph{CoRR}, abs/2208.10442.

\bibitem[{White et~al.(2023)White, Fu, Hays, Sandborn, Olea, Gilbert, Elnashar,
  Spencer-Smith, and Schmidt}]{White2023APP}
Jules White, Quchen Fu, Sam Hays, Michael Sandborn, Carlos Olea, Henry Gilbert,
  Ashraf Elnashar, Jesse Spencer-Smith, and Douglas~C. Schmidt. 2023.
\newblock A prompt pattern catalog to enhance prompt engineering with chatgpt.
\newblock \emph{ArXiv}.

\bibitem[{Wu et~al.(2023{\natexlab{a}})Wu, Bansal, Zhang, Wu, Zhang, Zhu, Li,
  Jiang, Zhang, and Wang}]{Wu2023AutoGenEN}
Qingyun Wu, Gagan Bansal, Jieyu Zhang, Yiran Wu, Shaokun Zhang, Erkang Zhu,
  Beibin Li, Li~Jiang, Xiaoyun Zhang, and Chi Wang. 2023{\natexlab{a}}.
\newblock Autogen: Enabling next-gen llm applications via multi-agent
  conversation framework.
\newblock \emph{ArXiv}.

\bibitem[{Wu et~al.(2023{\natexlab{b}})Wu, Irsoy, Lu, Dabravolski, Dredze,
  Gehrmann, Kambadur, Rosenberg, and Mann}]{Wu2023BloombergGPTAL}
Shijie Wu, Ozan Irsoy, Steven Lu, Vadim Dabravolski, Mark Dredze, Sebastian
  Gehrmann, Prabhanjan Kambadur, David Rosenberg, and Gideon Mann.
  2023{\natexlab{b}}.
\newblock Bloomberggpt: A large language model for finance.
\newblock \emph{ArXiv}.

\bibitem[{Yang et~al.(2020)Yang, Ng, Smyth, and Dong}]{Yang2020HTMLHT}
Linyi Yang, Tin Lok~James Ng, Barry Smyth, and Ruihai Dong. 2020.
\newblock Html: Hierarchical transformer-based multi-task learning for
  volatility prediction.
\newblock \emph{Proceedings of The Web Conference 2020}.

\bibitem[{Yoon et~al.(2022)Yoon, Woo, and Kim}]{Yoon2022HuBERTEEEE}
J.~Yoon, Beom~Jun Woo, and Nam~Soo Kim. 2022.
\newblock Hubert-ee: Early exiting hubert for efficient speech recognition.
\newblock \emph{ArXiv}.

\bibitem[{Yu and Li(2018)}]{Yu2018ForecastingSP}
Shui-Ling Yu and Zhe Li. 2018.
\newblock Forecasting stock price index volatility with lstm deep neural
  network.

\bibitem[{Zhang and Li(2023)}]{Zhang2023CanLL}
Yixuan Zhang and Haonan Li. 2023.
\newblock Can large langauge model comprehend ancient chinese? a preliminary
  test on aclue.
\newblock In \emph{International Conference on Algebraic and Logic
  Programming}.

\end{thebibliography}
